\documentclass{jps-cp}
\usepackage{txfonts}
\usepackage[colorlinks=true,
            linkcolor=blue,
            citecolor=blue,
            urlcolor=blue]{hyperref}
\makeatletter
\providecommand{\ext@figure}{lof}
\providecommand{\ext@table}{lot}
\makeatother
\usepackage{color}

\title{Towards Automatic Ion Identification in Storage Rings} 

\author{Anna Jule \textsc{Heß}$^{1}$, David \textsc{Freire-Fernández}$^{2,3,1}$, Moritz \textsc{Porstendörfer}$^{1}$, Wolfram \textsc{Korten}$^{2,3,4}$, and Yury A. Litvinov$^{1,3}$}

\inst{$^{1}$Institute for Nuclear Physics, University of Cologne, Cologne, Germany \\
$^{2}$FAIR Facility for Antiproton and Ion Research in Europe GmbH, 64291 Darmstadt, Germany
\\
$^{3}$GSI Helmholtzzentrum für Schwerionenforschung GmbH,
64291 Darmstadt, Germany\\
$^{4}$IRFU, CEA, Universit{\'e} Paris-Saclay, 91191 Gif-sur-Yvette, France}

\email{D.FreireFernandez@gsi.de}

\abst{
Resonant Schottky detectors provide non-destructive, single-ion-sensitive measurements {of revolution frequencies and intensities of ion species stored in a} heavy-ion storage ring. {These high-\(Q\) cavities allow analysis of high-order harmonics of the revolution frequencies to 
magnify} absolute frequency separations. Adjacent harmonic ranges can, however, overlap within the detector band, making frequency alone insufficient for {determining the ion's mass-to-charge ratio $(m/q)$}.
We present a proof-of-concept for an automated harmonic de-overlapping {method} using spectral shapes in combined Schottky and Isochronous Mass Spectrometry. 
The empirical ansatz is that, for a fixed ion-optical setting, each {peak, corresponding to a specific ion species with given $m/q$ ratio, has a}
characteristic profile that evolves 
{smoothly with changing $m/q$}.
Occurrences of the same local structure at different harmonics should therefore retain almost identical shapes after frequency rescaling. Applied to a 245\,MHz spectrum accumulated from 11\,452 injections at the Experimental Storage Ring at GSI, the method organizes 16 of 32 detected structures into five candidate families and reconstructs their fundamental-frequency representation. Reducing the complex mixed-harmonic spectrum to this smaller candidate set {
solves the challenging task of finding initial anchors for the} 
ion identification.}

\kword{Schottky detection, heavy-ion storage ring, isochronous mass spectrometry, harmonic de-overlapping, ion identification}

\begin{document}
\maketitle
\section{Introduction}

Non-destructive Schottky detection is central to mass, lifetime, and beam-diagnostic measurements with ions in storage rings \cite{Borer1974,Caspers2009, Litvinov2004,Franzke2008,SteckLitvinov2020,Sanjari2025}. A stored beam may contain tens of ionic species {with different intensities ranging from single to hundreds or even thousands of ions}. Successive passages of each ion through a {Schottky cavity} 
induce a periodic time-domain signal whose Fourier spectrum contains lines at
\begin{equation}
  f_h=h f_{\mathrm{rev}},
  \label{eq:harmonic}
\end{equation}
where $h$ is the harmonic number.

A resonant detector {has a maximum response within} a narrow band fixed primarily by its geometry and only slightly tunable \cite{Nolden2011,Sanjari2020}. Its high-$Q$ response enhances weak components. Moreover, for two signals observed at the same harmonic, their absolute separation scales as $\Delta f_h=h\Delta f_{\mathrm{rev}}$, which {is used to} reduce the observation time required to resolve them. Together, high sensitivity and increased absolute separation {achieved with the latest generation of Schottky cavity detectors\,\cite{Nolden2011,Sanjari2020}}, enable time-resolved measurements of individual stored ions in the millisecond regime~\cite{Freire2024,FreireThesis2024}.

\begin{figure}[t]
  \centering
  \begin{minipage}[t]{0.495\textwidth}
    \centering
    \includegraphics[width=\linewidth]{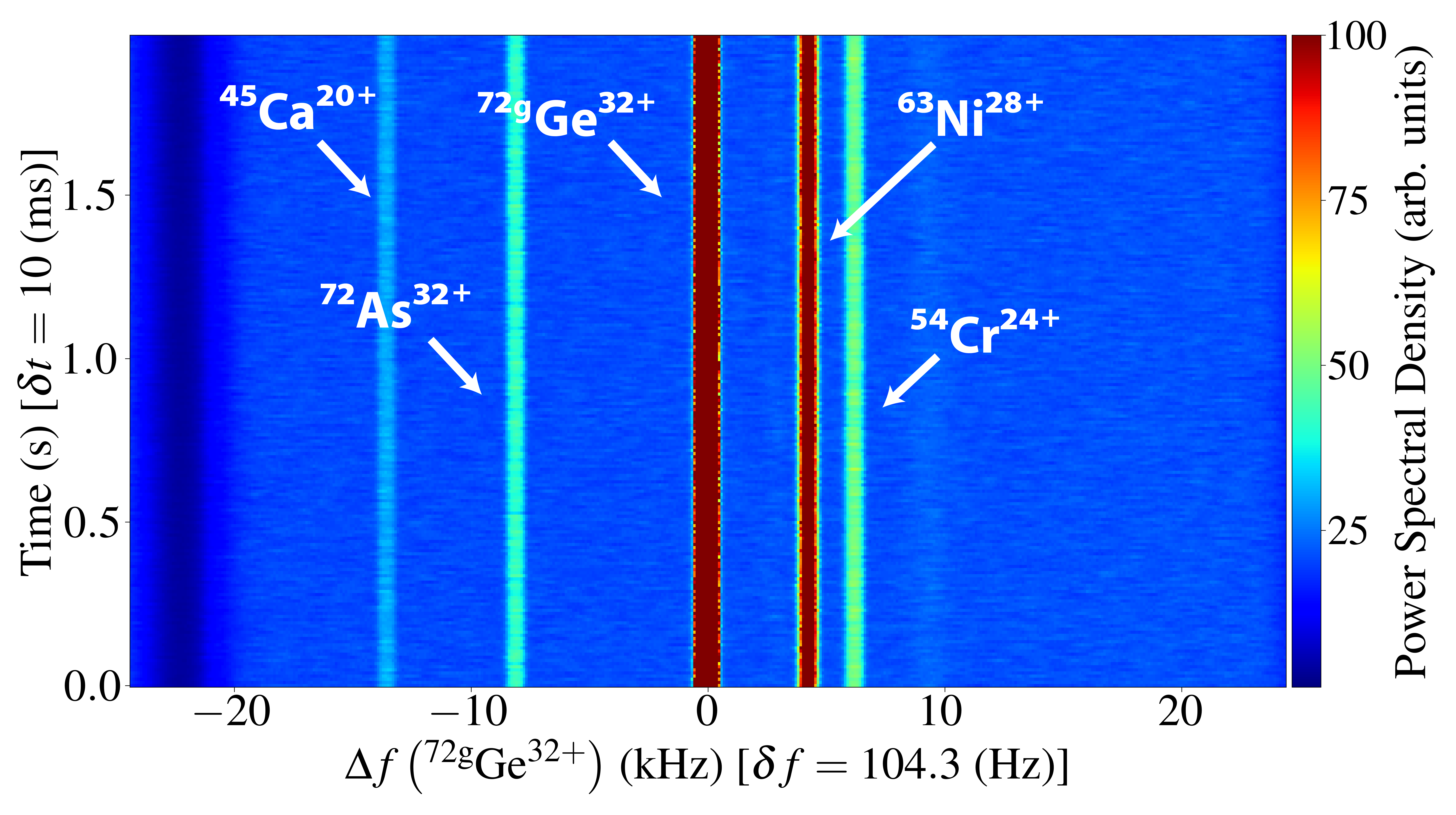}\\[-0.5ex]
    \textbf{(a)}
  \end{minipage}\hfill
  \begin{minipage}[t]{0.495\textwidth}
    \centering
    \includegraphics[width=\linewidth]{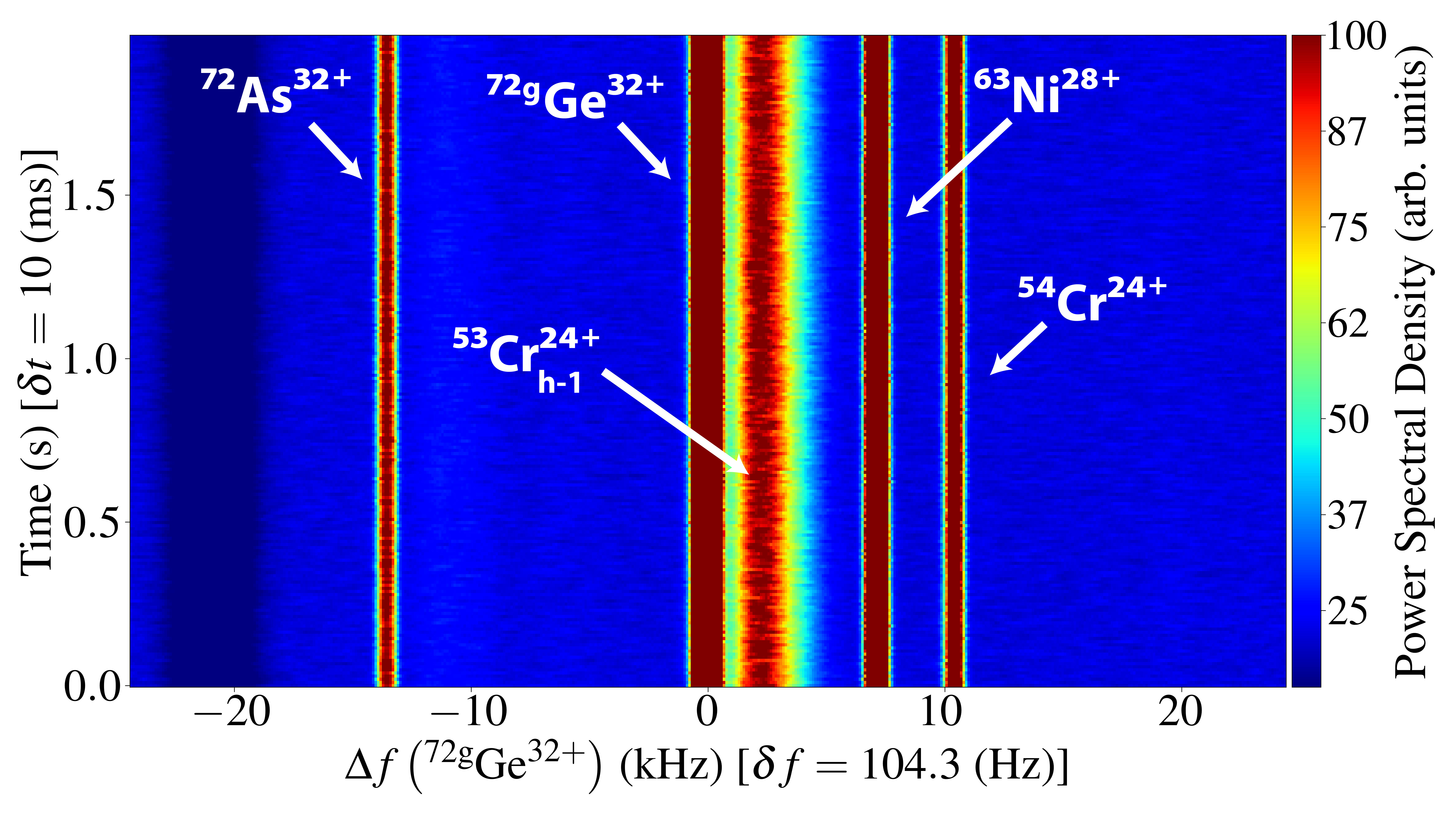}\\[-0.5ex]
    \textbf{(b)}
  \end{minipage}
  \caption{Schottky spectrograms of the same stored beam recorded with the ESR resonant detectors near (a) 245\,MHz\,\cite{Nolden2011} and (b) 410\,MHz\,\cite{Sanjari2020}. The additional broad $^{53}$Cr$^{24+}$ signal in (b) belongs to the adjacent lower harmonic, illustrating that nearby detected frequencies can correspond to different $m/q$. Ion labels were obtained with \textsc{RionID}~\cite{RionID}. Reproduced from Ref.~\cite{FreireThesis2024} under CC BY 4.0.}
  \label{fig:overlap}
\end{figure}

The same scaling expands the interval occupied by the complete ion distribution. Adjacent orders can therefore overlap, with $h_1 f_{\mathrm{rev},1}\simeq h_2 f_{\mathrm{rev},2}$ for $h_1\ne h_2$. 
{Within one harmonic, the variation of the revolution frequency directly reflects the variation of $m/q$\,\cite{Franzke2008}.
Hence, the assignment of all peaks to their correct harmonic numbers is decisive.}
The 245\,MHz {cavity}\,\cite{Nolden2011} 
observes approximately orders 120--130 for $f_{\mathrm{rev}}\simeq1.9$--$2.1$\,MHz. In Fig.~\ref{fig:overlap}, the broad $^{53}$Cr$^{24+}$ structure in the {band from} the 410\,MHz {cavity\,\cite{Sanjari2020}} belongs to an adjacent order, illustrating that frequency proximity does not imply proximity in $m/q$. Additionally, it can be seen how the distances between lines {of the same ion species} have increased considerably in Fig.~\ref{fig:overlap}(b) {as compared to Fig.~\ref{fig:overlap}(a)}.

The experiment used combined Schottky {plus} Isochronous Mass Spectrometry (S+IMS) \cite{Tu2018,Freire2024,Freire2026}. Unlike conventional Schottky mass spectrometry, the ions were not electron-cooled\,\cite{Radon00} {and} unlike conventional isochronous mass spectrometry, their repeated detection did not {utilize} destructive foil-based detection\,\cite{Hausmann00}. The measured line profiles therefore retain information about the revolution-frequency distributions produced by the injected phase-space distribution, the ring optics, and the frequency-dependent cavity response \cite{Freire2024,Sanjari2025,Freire2026}. Profiles close to the isochronous condition are {the most} narrow {ones} and approximately symmetric, whereas structures farther from it can broaden and develop asymmetric tails or shoulders.

We adopt the \emph{empirical ansatz} that, within a fixed setting, an ionic species with a given $m/q$ contributes a characteristic profile that changes {gradually} with $m/q$. Nearby components form local structures that can recur at successive harmonics. Such occurrences probe the same underlying revolution-frequency distribution and should retain related profiles after frequency rescaling and normalization, although the cavity response, finite resolution, and overlapping signals can deform them. Shape information is therefore used only together with the integer periodicity of Eq.~(\ref{eq:harmonic}). Mapping the resulting candidate families to $h=1$ reduces the assignment space and, therefore, allows {the ion identification software} \textsc{RionID} to optimize ring and calibration parameters over a much smaller set of possibilities---a first step toward automatic ion identification.

\section{Data and reconstruction procedure}

{Used here is a subset of data acquired in the} {FAIR Phase-0} experiment G-22-00203 at the {GSI Experimental Storage Ring (ESR)} 
Projectile fragments from an approximately 512\,MeV/u $^{208}$Pb beam incident on a 10-mm-thick $^{9}$Be target were injected into the isochronously operated ring. The 245\,MHz {cavity} 
was read out with a real-time spectrum analyzer. The dataset contains 11\,452 injections {each} of approximately 3\,s duration. Their {Fast Fourier Transform} power spectra were {summed together,}
and the slowly varying instrumental baseline was subtracted following {the procedure described in} Ref.~\citen{Wang2022}. 

\begin{figure}
    \centering
    \includegraphics[width=0.75\linewidth]
        {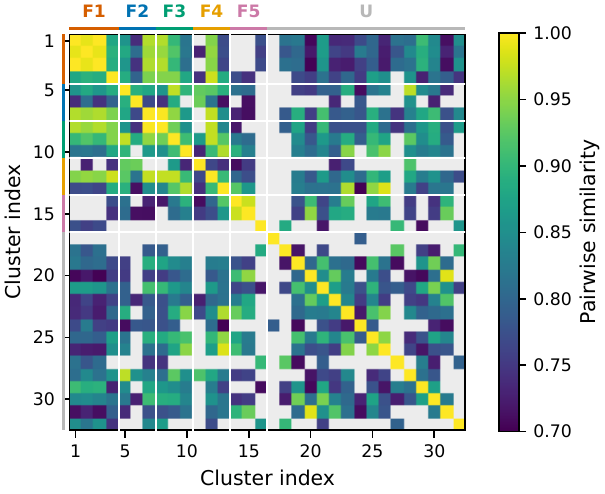}
    \caption{Pairwise cosine similarity of the 32 normalized cluster templates. Entries below $S_{ij}^{\mathrm{shape}}=0.70$ are masked. Colored strips identify the candidate harmonic families F1--F5;
    U denotes clusters that were not assigned to an accepted family. Clusters are ordered by their family assignment rather than by frequency.}
    \label{fig:similarity_matrix}
\end{figure}

The reconstruction consists of four stages:

\noindent\textit{(1) Peak detection and clustering.} Peaks are detected using prominence in the cavity-resonance region and a local threshold elsewhere. Neighbors are clustered according to separation and relative amplitude. A cluster may contain one peak or several partially resolved components.

\noindent\textit{(2) Shape representation.} The constituent profiles of cluster $i$ are mapped to a common coordinate and $L^2$-normalized. Their normalized mean defines the template $\boldsymbol{T}_i$, with $\lVert\boldsymbol{T}_i\rVert_2=1$. Pairwise similarity is
\begin{equation}
  S_{ij}^{\mathrm{shape}}
  =\boldsymbol{T}_i\!\cdot\!\boldsymbol{T}_j,
  \label{eq:cosine}
\end{equation}
{for} identical templates $S_{ij}^{\mathrm{shape}}=1$.

\noindent\textit{(3) Periodic grouping and ranking.} Shape-compatible clusters are tested as harmonic families. Integer labels $h_i$ are assigned within $f_{\mathrm{rev}}=2.0\pm0.1$\,MHz, and the measured positions are fitted with
\begin{equation}
 f_i=f_{\mathrm{off}}+h_i f_{\mathrm{rev}}.
 \label{eq:fit}
\end{equation}
For Eq.~(\ref{eq:harmonic}), $f_{\mathrm{off}}=0$; a free intercept serves as a diagnostic for calibration or assignment biases. A family must contain at least three occurrences and use each harmonic at most once, favoring strong structures visible over several orders. Candidates are ranked by
\begin{align}
S_{\mathrm{family}}={}&0.30S_{\mathrm{freq}}+0.15S_{\mathrm{shape}}
+0.10S_{\mathrm{amp}}+0.05S_{\mathrm{res}}\notag\\[-1mm]
&+0.05S_{\mathrm{size}}+0.35S_{\mathrm{order}},
\label{eq:score}
\end{align}
which combines frequency residuals, shape and amplitude consistency, cavity-response position, family size, and harmonic ordering. This empirical score is a heuristic ranking, not a statistical confidence. Conflicts are resolved in favor of higher-scoring, non-overlapping families.

\noindent\textit{(4) Folding.} A linear fit of Eq.~(\ref{eq:fit}) yields $f_{\mathrm{rev}}$ as its slope for each accepted family. The assigned integer $h_i$ then maps the corresponding profile to $h=1$ through $f_{\mathrm{folded}}=f/h_i$. Folded profiles are interpolated onto a common grid and averaged. The resulting candidate spectrum is passed to \textsc{RionID}~\cite{RionID} for assignment.

\section{Results for the 245\,MHz spectrum}

The peak search produced 32 clusters. Figure~\ref{fig:similarity_matrix} shows {cases of} substantial off-diagonal similarity, including clusters which are not {assigned to any} family, while accepted members need not to be identical. Shape therefore supports the grouping but cannot determine it alone.

\begin{figure}
    \centering
    \includegraphics[width=\linewidth]
        {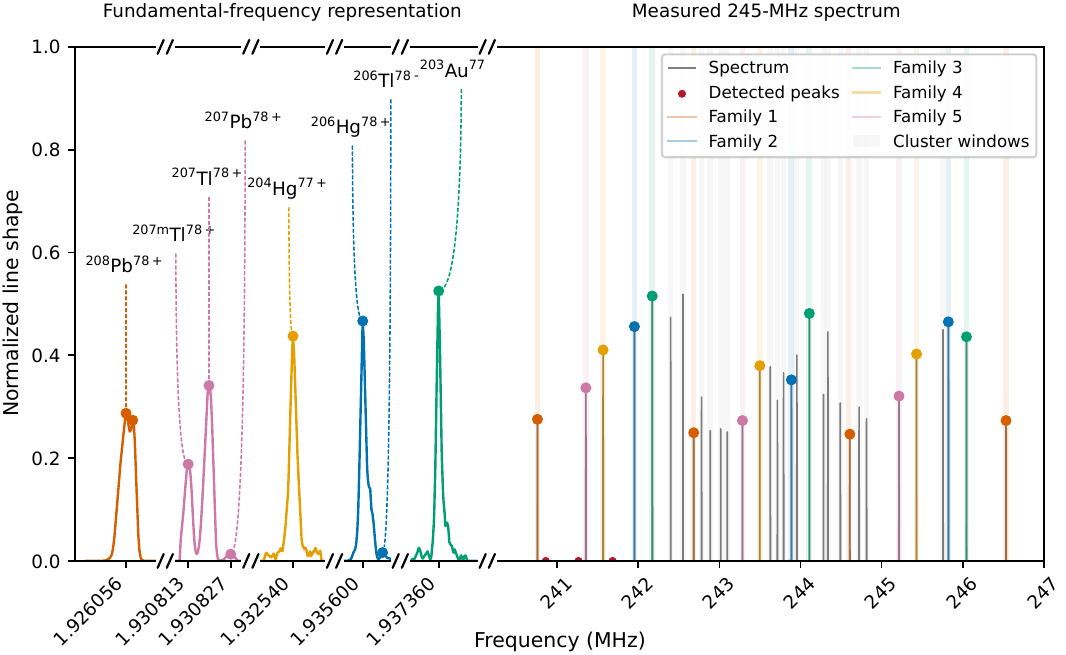}
    \caption{Harmonic-family reconstruction. Left: profiles folded to $h=1$ with the assigned harmonic orders and averaged within each family; internal breaks compress the gaps between families. The central break separates the folded and measured-frequency axes. Right: normalized accumulated 245\,MHz spectrum. Dots mark detected peaks, shaded bands show cluster locations, and colors identify the candidate harmonic families F1--F5. {Peak} widths are enlarged for visibility. Ion labels are {results of the} \textsc{RionID} {identification}.}
    \label{fig:frequency_folding}
\end{figure}

Five families passed the combined criteria and contain 16 occurrences: four in F1 and three in each of F2--F5, see Fig.\,\ref{fig:similarity_matrix}. Their reconstructed fundamental frequencies are 1.9261, 1.9308, 1.9325, 1.9356, and 1.9374\,MHz, {respectively,} with assigned orders near $h=125$--128. Figure~\ref{fig:frequency_folding} connects their measured locations to the folded $h=1$ representation.

The region near 1.9261\,MHz is compatible in \textsc{RionID} with $^{208}$Pb$^{78+}$. Near 1.9308\,MHz, the candidate components are $^{207\mathrm{m}}$Tl$^{78+}$, $^{207}$Tl$^{78+}$, and $^{207}$Pb$^{78+}$. Further matches are $^{204}$Hg$^{77+}$ near 1.9325\,MHz, $^{206}$Hg$^{78+}$ and $^{206}$Tl$^{78+}$ near 1.9356\,MHz, and $^{203}$Au$^{77+}$ near 1.9374\,MHz. Thus, {the} recovered family can represent a recurring local multi-component structure rather than one isolated peak. 

\section{Conclusion and outlook}

We have demonstrated a proof-of-concept procedure that combines local spectral-shape similarity {analysis} with the integer periodicity of Schottky harmonics. From 32 clusters in the accumulated {spectrum from the 245\,MHz resonator}, the procedure retained 16 occurrences in five candidate families and mapped them from harmonics near 125--128 to a compact $h=1$ representation around 1.93\,MHz. The reconstructed positions are compatible with plausible \textsc{RionID} assignments. The principal result is therefore a reduction of a complex, mixed-harmonic spectrum to a much smaller and physically constrained assignment problem. 

The immediate next step is to connect the reconstructed $h=1$ spectrum directly to \textsc{RionID}, allowing ring, calibration, charge-state, and harmonic assignments to be automatically optimized jointly. A joint treatment of the 245 and 410\,MHz detectors would enable identifying weak signals that are only present in the resonance regions. Finally, template and similarity uncertainties should be propagated into the ranking, and the empirical weights in Eq.~(\ref{eq:score}) further optimized. These steps could turn the present offline proof of concept into a reproducible tool for online ion identification {during} future heavy-ion storage-ring experiments {at GSI and elsewhere\,\cite{Yamaguchi, 153Hf}}.

\section*{Acknowledgements}
{The methods presented here are being prepared for the analysis of 
the joint experiment G-22-00203 and G-22-00018, which was performed at the ESR
storage ring of the GSI Helmholtzzentrum f{\"u}r
Schwerionenforschung, Darmstadt (Germany) in the context of FAIR Phase-0 research program.}

{\footnotesize


\begin{thebibliography}{99}
\setlength{\itemsep}{0pt}
\setlength{\parskip}{0pt}

\bibitem{Borer1974}
J. Borer, P. Bramham, H. G. Hereward, K. H\"ubner, W. Schnell,
and L. Thorndahl,
in \textit{Proc. IXth International Conference on High Energy
Accelerators}, Stanford, USA, 1974
(SLAC, Stanford, 1974), pp. 53--56.

\bibitem{Caspers2009}
F. Caspers,
in \textit{Proc. CERN Accelerator School: Course on Beam Diagnostics},
Dourdan, France, 2008,
CERN-2009-005
(CERN, Geneva, 2009), pp. 407--425.

\bibitem{Litvinov2004}
Yu. A. Litvinov, H. Geissel, Yu. N. Novikov, \textit{et al.},
Nucl. Phys. A \textbf{734}, 473 (2004).
\bibitem{Franzke2008}
B. Franzke, H. Geissel, and G. M\"unzenberg,
Mass Spectrom. Rev. \textbf{27}, 428 (2008).

\bibitem{SteckLitvinov2020}
M. Steck and Yu. A. Litvinov,
Prog. Part. Nucl. Phys. \textbf{115}, 103811 (2020).

\bibitem{Sanjari2025}
M. S. Sanjari \textit{et al.},
Chin. Phys. C \textbf{49}, 084007 (2025).

\bibitem{Nolden2011}
F. Nolden, P. H\"ulsmann, Yu. A. Litvinov, \textit{et al.},
Nucl. Instrum. Methods Phys. Res., Sect. A \textbf{659}, 69 (2011).

\bibitem{Sanjari2020}
M. S. Sanjari \textit{et al.},
Rev. Sci. Instrum. \textbf{91}, 083303 (2020).

\bibitem{FreireThesis2024}
D. Freire-Fern\'andez,
Ph.D. thesis, Heidelberg University (2024), doi:10.11588/heidok.00035046.

\bibitem{Freire2024}
D. Freire-Fern\'andez \textit{et al.},
Phys. Rev. Lett. \textbf{133}, 022502 (2024).

\bibitem{Tu2018}
X. L. Tu \textit{et al.},
Phys. Rev. C \textbf{97}, 014321 (2018).

\bibitem{Freire2026}
D. Freire-Fern\'andez \textit{et al.},
arXiv:2607.10894 [nucl-ex] (2026).

\bibitem{Wang2022}
Q. Wang, X. Yan, X. Chen, P. Shuai, M. Wang, and Y.-H. Zhang,
Nucl. Sci. Tech. \textbf{33}, 148 (2022).

\bibitem{RionID}
D. Freire-Fern\'andez, G. Hudson-Chang, and R.-J. Chen,
\textsc{RionID}: Ring-stored ion IDentification, version 9.0.0,
Zenodo (2026), doi:10.5281/zenodo.22041793.

\bibitem{Radon00}
T.~Radon \textit{et al.}, Nucl. Phys. A \textbf{677}, 75 (2000).

\bibitem{Hausmann00}
M. Hausmann \textit{et al.}, Nucl. Inst. Meth. A \textbf{446}, 569 (2000).

\bibitem{Yamaguchi}
T. Yamaguchi \textit{et al.},
Prog. Part. Nucl. Phys. \textbf{120}, 103882 (2021).

\bibitem{153Hf}
H. F. Li \textit{et al.},
Sci. Bull. \textbf{71}, 4099 (2026).

\end{thebibliography}
\end{document}